\documentclass[10pt,conference]{IEEEtran}
\IEEEoverridecommandlockouts
\usepackage{cite}
\usepackage{amsmath,amssymb,amsfonts}
\usepackage{algorithmic}
\usepackage{graphicx}
\usepackage{tabularx, booktabs}
\usepackage{textcomp}
\usepackage{xcolor}
\usepackage{footnote}
\makesavenoteenv{tabular}
\makesavenoteenv{table}

\newcommand{\multilinecell}[2][c]{%
  \begin{tabular}[#1]{@{}l@{}}#2 \end{tabular}}

\def\BibTeX{{\rm B\kern-.05em{\sc i\kern-.025em b}\kern-.08em
    T\kern-.1667em\lower.7ex\hbox{E}\kern-.125emX}}

\makeatletter
\newcommand{\linebreakand}{%
  \end{@IEEEauthorhalign}
  \hfill\mbox{}\par
  \mbox{}\hfill\begin{@IEEEauthorhalign}
}
\makeatother

\begin{document}

\title{A Network Perspective on the Influence of Code Review Bots on the Structure of
Developer Collaborations*\thanks{* Accepted at the CHASE 2023 Registered Reports Track.}
}

\DeclareRobustCommand*{\IEEEauthorrefmark}[1]{%
  \raisebox{0pt}[0pt][0pt]{\textsuperscript{\footnotesize #1}}%
}

\author{
\IEEEauthorblockN{Leonore Röseler\IEEEauthorrefmark{1}\textsuperscript{,}\IEEEauthorrefmark{2}}
\IEEEauthorblockA{leonore@ifi.uzh.ch}
\and
\IEEEauthorblockN{Ingo Scholtes\IEEEauthorrefmark{1}\textsuperscript{,}\IEEEauthorrefmark{3}}
\IEEEauthorblockA{ingo.scholtes@uni-wuerzburg.de}
\and
\IEEEauthorblockN{Christoph Gote\IEEEauthorrefmark{1}\textsuperscript{,}\IEEEauthorrefmark{4}}
\IEEEauthorblockA{cgote@ethz.ch}
\linebreakand
\IEEEauthorblockA{\IEEEauthorrefmark{1}\textit{Data Analytics Group}, University of Zurich, Zurich, Switzerland}
\IEEEauthorblockA{\IEEEauthorrefmark{2}\textit{Social Computing Group}, University of Zurich, Zurich, Switzerland}
\IEEEauthorblockA{\IEEEauthorrefmark{3}\textit{Chair for Machine Learning for Complex Networks}, Julius-Maximilians-Universität Würzburg, Würzburg, Germany}
 \IEEEauthorblockA{\IEEEauthorrefmark{4}\textit{Chair of Systems Design}, ETH Zurich, Zurich Switzerland}
}

\maketitle

\begin{abstract}
\textbf{Background:} Despite a growing body of literature on the impact of software bots on open source software development teams, their effects on team communication, coordination, and collaboration practices are not well understood.
Bots can have negative consequences, such as producing information overload or reducing interactions between developers.

\textbf{Objective:} The objective of this study is to investigate the effect of specific GitHub Actions on the collaboration networks of Open Source Software teams, using a network-analytic approach.
The study will focus on Code Review bots, one of the most frequently implemented types of bots.

\textbf{Method:} Fine-grained, time-stamped data of co-editing networks, developer-file contribution networks, as well as workflow runs and git commit logs will be obtained from a large sample of GitHub repositories.
This will allow us to study how bots affect the collaboration networks of developers over time.
By using a more representative sample of GitHub repositories than previous studies, which includes projects whose sizes span the whole range of Open Source communities, this study will provide generalizable results and updated findings on the general usage and distribution of GitHub Actions.
With this study, we aim to contribute to advancing our knowledge of human-bot interaction and the effects of support tools on software engineering teams.

\end{abstract}


\section{Introduction}

\subsection{Context}

Social coding is an approach that promotes collaboration among developers and is generally performed through a platform for distributed and asynchronous collaboration, or supported through other tools \cite{dabbishSocialCodingGitHub2012, schunemannSocialCodingRevolution2018}.
Bots and continuous integration tools are used to support development processes through the automation of workflows. The \verb|TODO to Issue| bot, for example, automatically converts newly committed \verb|TODO| comments to GitHub issues in response to a push event.
Bots designed to help developers become more productive entered the stage long ago, but have experienced an upswing recently \cite{lebeufDefiningClassifyingSoftware2019}.
In October 2018, GitHub -- one of the most prominent social coding platforms -- introduced the so-called GitHub Actions (GAs) 
in a beta version \cite{GitHubActionsFocus2023}. These were made available for public use only one year later \cite{GitHubActionsGenerally2023}.
GAs are customizable bot programs for automating workflows and supporting team coordination. 
The \verb+Pull Request size labeler+ bot, for example, is supposed to help developers better grasp the size of a pull request (PR) and limit its size if desired.

Although software development bots are not new, they are more frequently integrated into the development process and with the improvement of AI technology, they have new potential to support developers; see also \cite{erlenhovCurrentFutureBots2019, storeyDisruptingDeveloperProductivity2016, lebeufDefiningClassifyingSoftware2019, zampettiEmpiricalCharacterizationBad2020}.

The inclusion of bots in the development process can be beneficial:
By automating workflows, developers' resources can be freed up so that developers can be more
productive by focusing on other tasks. According to a survey by Rue, 37\% of developers in
the US spend more than 25\% of their time fixing bugs \cite{rueStateSoftwareCode2021}. 
Thus, especially bots for code review and bug detection can benefit projects
monetary-wise. Furthermore, bots can help by optimally matching tasks and developers, which
enhances team coordination, and by raising awareness of open issues, for example. 
There are indications that bot services benefit computer science students
\cite{huImprovingFeedbackGitHub2019}  and that there is a positive perception
of GAs among developers, in general \cite{kinsmanHowSoftwareDevelopers2021a}.

On the other hand, bots can also bring challenges, such as a reduction of team interactions and
thereby eliminate chances for serendipitous learning
\cite{storeyDisruptingDeveloperProductivity2016}. In their survey, Wessel et al.
found that ``almost 50\% of the contributors and more than 50\% of the integrators
disagreed with the premise that bots improve social interaction''
\cite{wesselPowerBotsUnderstanding2018}. 
In their study on the effects of Code Review bots on Open Source Software (OSS) project
activity, Wessel et al. suggested that new technology, such as bots for OSS projects
``may bring unexpected impacts to group dynamics''
\cite{wesselQualityGatekeepersInvestigating2022}. For instance, they found an indication for a decrease
in communication among developers (where communication was defined as the number of
comments on (non)merged pull requests (PRs)).
Farah et al. \cite{farahExploratoryStudyReactions2022a} observed that bot comments receive
more \verb+-1+ comments than humans, that is, negative feedback, suggesting that users
perceive bots more negatively or are more willing to give them negative feedback.
Especially \verb+stale[bot]+, a bot responsible for closing stale issues, received
disproportionate negative feedback.

Due to recent rapid improvements of artificial intelligence, more and more tools are AI-powered. 
After the recent hype around the launch of ChatGPT, a chatbot developed by OpenAI based on a large language model, developers started to make an attempt to integrate ChatGPT into an Action on the GitHub Actions marketplace.
Another example of how to leverage AI for code review is the Action Codeball \cite{CodeballAICode2023}.
With the rise of AI, it is an open question how GAs such as Codeball influence
how software engineering teams develop and collaborate.

\subsection{Problem Statement}

While some studies have investigated the more general impact of continuous integration tools and software bots on open source development practices, the effects of GAs in general and specific types of GAs on the structure and dynamics of developer collaborations are not well understood yet.
It remains an open question as to how the adoption of certain bots on GitHub affects the way developers communicate, coordinate, and collaborate.
It is also unclear how we can prevent or at least mitigate the potentially negative effects of bots, such as increased information overload and distraction \cite{wesselBotsPullRequests2022}, or a potential decrease in social interactions that could damage the social cohesion of teams \cite{deyImpactTeamDesign2020}.

Previous studies primarily looked at changes of certain activity metrics, such as the number of (non)merged PRs or the time to close PRs, in an isolated way, aggregated at the project level for monthly units \cite{wesselQualityGatekeepersInvestigating2022, wesselGitHubActionsImpact2022,chenLetSuperchargeWorkflows2021,kinsmanHowSoftwareDevelopers2021a}, and did not investigate how the complex interactions of individual developers within a team change with the adoption of bots.  

We believe that a network-analytic perspective that takes into account team interactions and how developers \textit{co-edit} will produce a better understanding of the effects of GAs and will also help illuminate the somewhat mixed findings of prior research.

\subsection{Goal}

Addressing the research gap outlined above, the goal of our study is to investigate the effect of specific GAs on the collaboration structures of developers in OSS teams. To this end, we will employ network-analytic techniques that leverage time-stamped interactions automatically mines from git repositories at large scale.

We will focus on the effects of Code Review bots, which will enable us to compare our findings with those of a previous study
\cite{wesselQualityGatekeepersInvestigating2022}.
The rationale behind this specific choice is that Code Review bots are one of the most frequently adopted type of bots.
Moreover, they have the potential to affect how developers co-edit each other's code.
By taking into account the timestamps of when workflows were run as well as fine-grained, time-stamped data on co-edits generated in a large sample of GitHub repositories, we will have the opportunity to associate bot effects with detailed data on development actions with high temporal resolution.

\subsection{Contributions}

Studies on the effects of bot adoption in software engineering teams so far are often based on the analysis of aggregated data from different GAs, using a regression discontinuity design to investigate changes in activity indicators after bot adoption \cite{chenLetSuperchargeWorkflows2021,wesselGitHubActionsImpact2022,wesselQualityGatekeepersInvestigating2022, kinsmanHowSoftwareDevelopers2021a}. 
Most of these studies focused on highly popular projects
\cite{chenLetSuperchargeWorkflows2021,wesselGitHubActionsImpact2022}, which bears the risk of producing biased findings that do not generalize to smaller development teams, which, however, constitute the vast majority of projects on popular social coding platforms such as GitHub \cite{goteBigDataBig2022}.
Instead, we plan to obtain a more representative sample of OSS repositories compared to previous studies by utilizing a sample that also includes less popular projects.

Moreover, our study will be the first to use advanced network-analytic methods to study whether and how code review bots change the structure of interactions between developers, i.e., \textit{collaboration networks} in a repository.
Especially, we will investigate whether bots are associated in a statistically significant way with differences in the topology of networks that capture which developers co-edit each other's code.

To investigate the impact of bots, we will take a two-fold approach: 
We first perform a statistical comparison of time-aggregated data on projects with and without bots. 
The additional inclusion of the temporal dimension of our data will enable us to take one step further towards a causal interpretation of our findings (notwithstanding the fact that a strictly causal interpretation will be challenging). 
To this end, we will conduct a temporal comparison for projects adopting code review bots, examining whether topological metrics of collaboration networks differ significantly before and after bot adoption.

Through this new methodological approach, our aim is to gain new insights into the impact of (Code Review) bots on OSS teams by analyzing changes in the collaboration networks of developers and also replicating some of the observations from earlier studies.
For example, our study will produce updated findings on the general usage and distribution of GAs across repositories. 
Considering also less popular projects, our sample will be more representative of the general population of (public) projects on GitHub.
As the use of GAs has probably still increased since their official launch in 2019 \cite{GitHubActionsGenerally2023}, we expect to observe both a higher number and a more varied selection of GAs.
For our main analysis, Code Review bots will be our focus of attention, for which we will be able to replicate findings from a study by Wessel et al. \cite{wesselQualityGatekeepersInvestigating2022}.
An interesting new aspect will be the inclusion of AI-based Code Review bots, such as Codeball, in the sample of Code Review bots, which may lead to different changes in the development practices compared to previously studied bots.

With this study, our aim is to contribute to advancing our knowledge of human-bot interaction and support tools for software engineering teams, which may even be AI-powered.
We aim to extend prior findings through a new methodological approach, that is, statistical network analysis of collaboration networks. 
Findings about how bots cause changes in collaboration network metrics may help us to make inferences about how bots influence software development. 
But findings that suggest an impact of bots may also ``just'' inform research that relies on data originating from version control systems, which are affected by bots.

\section{Background and Related Work}

\subsection{GitHub Actions}

Since the introduction of GAs, they have become a popular service used in GitHub OSS repositories, often replacing external services such as Travis CI \cite{golzadehRiseFallCI2022}, a tool for continuous integration. 
GitHub offers templates for GAs and allows developers to highly customize workflows.
The use of GAs is widely spread \cite{golzadehRiseFallCI2022}, especially in larger projects.
To the best of our knowledge, there does not exist a study yet that fully investigated the use of GAs in the \textit{total population} of GitHub repositories and the association of project characteristics with GA adoption.
While numerous papers on GitHub Actions exist, we are not aware of a study based on a sample drawn from the total population of GitHub repositories, that is, based on a sample that also  comprises less popular and smaller repositories. 
Rather, we observed substantial variability in how projects were selected; compare also 
\cite{golzadehRiseFallCI2022}, 
\cite{wesselBotsPullRequests2022} and
\cite{wesselPowerBotsUnderstanding2018}.
It can be assumed that many projects use GAs to automate their workflows.
For example, Wessel et al. observed that about 30\% of the projects adopted at least one GA, based on a sample of 5,000 most-starred projects on GitHub \cite{wesselGitHubActionsImpact2022}. 
Chen et al. found that 22\% of popular projects, i.e., projects with more than 1,000 stars, adopted GAs and that this adoption was correlated with project contributor size \cite{chenLetSuperchargeWorkflows2021}. In their study, each project adopted 2.8 GAs on average.

Several studies investigated the effects of continuous integration or bot adoption on development practices, looking at changes in activity indicators over time.
Studies focused on the adoption of the Travis Continuous Integration tool \cite{zhaoImpactContinuousIntegration2017}, GAs \cite{kinsmanHowSoftwareDevelopers2021a, chenLetSuperchargeWorkflows2021, golzadehRiseFallCI2022, wesselGitHubActionsImpact2022}, and a variety of bots \cite{wesselPowerBotsUnderstanding2018} or a specific type of bot \cite{wesselQualityGatekeepersInvestigating2022, moharilJIRAGitHubASFBot2022, mohayejiAdoptionTODOBot2022}. 
Findings are mixed and seem to depend on the purpose of a bot; see also \cite{wesselGitHubActionsImpact2022}:

In an early study based on about 350 popular GitHub projects of which about a fourth employed some kind of bot, researchers did not find significant differences for the time before and after bot adoption in the number of comments, number of commits, number of changed files, and time to close PRs. \cite{wesselPowerBotsUnderstanding2018}. 

Chen et al. observed that after GA adoption the commit frequency, the number of closed issues, as well as the time to close issues decreased \cite{chenLetSuperchargeWorkflows2021}.
Additionally, the number of PRs seemed to decrease and
the PR resolution latency decreased.
In another study, Kinsman et al. observed more rejected PRs and fewer commits to merged PRs after GA adoption \cite{kinsmanHowSoftwareDevelopers2021a}.

Focusing on the adoption of Travis CI, Zhao et al. found that the number of merge commits increased, while the guideline to make small commits was only followed to some extent, with large differences between projects \cite{zhaoImpactContinuousIntegration2017}.
Furthermore, the number of closed pull requests increased, the pull request latency increased despite the code changes becoming smaller, and the trend of increasing closed issues slowed down.

Focusing on a specific type of bots implemented in about 1,200 projects, Code Review bots such as Codecov, Wessel et al. \cite{wesselQualityGatekeepersInvestigating2022} observed an increased number of merged PRs and decreasing number of unmerged PRs, which speaks for the bot pushing contributors to enhance their PRs.
The authors also state that they observed decreasing communication among developers in terms of comments on merged PRs. 

Investigating the effects of the \verb+TODO+ bot, which automatically creates a GitHub issue when a
\verb+TODO+ comment is written in the code, Mohayeji et al. unsurprisingly found an increase in
\verb+TODO+ comments.
However, despite the fact that contributors increasingly used the bot, there was no speed-up in \textit{addressing} the open issues
\cite{mohayejiAdoptionTODOBot2022}.
As this example shows, a bot may not be as effective as one might
hope.

To investigate the effect bot adoption has on the change in activity indicators, all related studies used regression discontinuity design and mostly aggregated data on a monthly basis \cite{zhaoImpactContinuousIntegration2017, moharilJIRAGitHubASFBot2022, casseeSilentHelperImpact2020, wesselQualityGatekeepersInvestigating2022, guoStudyingImpactCI2019, chenLetSuperchargeWorkflows2021, zimmermannImpactSwitchingBug2019, kinsmanHowSoftwareDevelopers2021a}.
In our work, we will study the impact of bots on communication, coordination, and collaboration in OSS projects by looking at changes in the topology of collaboration networks. 
Leveraging techniques from social and statistical network analysis to investigate developer networks has often been proven successful \cite{birdDonTouchMy2011, birdPuttingItAll2009a, joblinHowSuccessfulFailed2022, joblinDeveloperNetworksVerified2015, meneelySociotechnicalDeveloperNetworks2011, lopez-fernandezApplyingSocialNetwork2006}.

\section{Research Questions}

We assume that Code Review bots impact how developers collaborate and that this does not only show in changes of activity indicators aggregated for a whole project, as previous studies did \cite{wesselQualityGatekeepersInvestigating2022, wesselGitHubActionsImpact2022,chenLetSuperchargeWorkflows2021,kinsmanHowSoftwareDevelopers2021a}.
But, that this also becomes apparent in changes in the interactions between individual developers.
For example, bot support could have the effect that developers need to coordinate less, which could become evident through changes in co-editing networks.

In this exploratory study, we address two research questions:

\textbf{RQ1:} Does the adoption of Code Review bots influence the structure of \textbf{collaboration networks} between developers in Open Source software projects?

To answer this question, we will adopt a temporal network analysis perspective that utilizes time-stamped co-editing networks of developers, which can be automatically generated based on the Open Source package \textit{git2net} \cite{goteGit2netMiningTimeStamped2019}.

This package uses the commit log history of git repositories to infer co-editing relationships. This means that a time-stamped co-editing relationship between developer A and developer B is inferred if developer A edited a line of code that was previously written by developer B.
Previous studies have shown that the resulting time-stamped co-editing networks carry rich information on the evolving collaboration structures in software projects \cite{goteBigDataBig2022, trujilloCorrectionPenumbraOpen2022, goteAnalysingTimeStampedCoEditing2021a, scholtesAristotleRingelmannLargeScale2016}.
To answer RQ1, we propose to test the following hypotheses.

\textbf{H1:} The adoption of Code Review bots is associated with differences in the structure of developer collaborations in Open Source software projects.

\textbf{H2:} The adoption of Code Review bots in Open Source projects precedes the change of collaboration structures.

\label{test-h1}
To test H1, we construct two representative samples of GitHub projects, one consisting of projects adopting a Code Review bot and one consisting of projects not using such a bot.
We will sample projects without bots in such a way that these projects are similar to projects with bots w.r.t. their general characteristics. 
For each of the two samples, using \textit{git2net}, we will generate a single time-aggregated weighted co-editing network, for which we will calculate pre-determined network-level metrics.
Subsequently, we will conduct inter-project comparisons. More specifically,  
we will test whether significant differences between the two samples w.r.t. the calculated network metrics exist.

Importantly, even if we find significant differences between the two groups of projects, this does not necessarily imply a causal relationship in the sense that the adoption of Code Review bots causes changes in the topology of co-editing networks.
For example, it is conceivable that projects which suffer from a lack of collaboration among developers are more likely to adopt Code Review bots.
To address this problem, we will use a temporal perspective and additionally test the second hypothesis, H2, that the adoption of Code Review bots \emph{temporally precedes} the change of collaboration structures. 
\label{test-h2}
To test H2, we conduct intra-project comparisons for the projects adopting a Code Review bot.
Assuming that bot adoption impacts collaboration structures, we expect to observe significant changes in network metrics between the period before (phase 1) and after bot adoption (phase 2). We determine the phases to span six months, respectively, following prior literature \cite{wesselGitHubActionsImpact2022, wesselPowerBotsUnderstanding2018}.
Again, for each project, we will generate time-aggregated co-editing networks, one for phase 1 and one for phase 2. Then, we will test the hypothesis that there is a significant change in the collaboration structures between the two phases, measured as a significant change in the network metrics.

To control for potential effects of project aging or other temporal effects, we will conduct multiple intra-project comparisons, based on randomly drawn, additional time-points which do not fall into the time interval of bot adoption, i.e., neither into phase 1 nor phase 2.
For these additional time-points, we do not expect to observe changes in the network metrics as large as the changes between phase 1 and phase 2.

\begin{figure}[htbp]
    \centerline{\includegraphics[width=\linewidth]{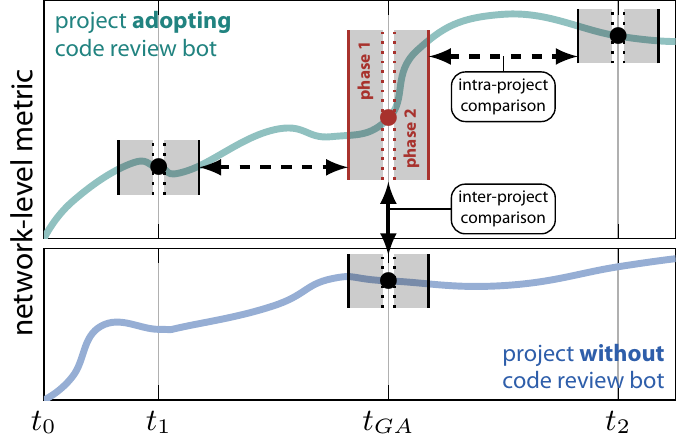}}
    \caption{Illustration of study design with inter-project comparison and intra-project comparisons. $t_0$ denotes the start of a project, operationalized by the first commit. Projects with and without bots are matched to have (approximately) equal project age. The green line in the top row depicts an
    exemplary development of some network-level metric over time for a project with a bot. 
    For projects with a bot, $t_{GA}$ denotes the time-point at which a Code review bot was introduced first; for projects without a bot, it denotes the corresponding time-point in the project life-cycle. 
    The blue line in the bottom row depicts an exemplary development of the network-level metric over time for a project without a bot.
    $t_1$ and $t_2$ denote randomly drawn time-points in the life-cycle of projects, outside of phase 1 and phase 2.}
    \label{fig:study-design}
\end{figure}

Both inter-project and intra-project comparisons are illustrated in Figure \ref{fig:study-design}.
The infographic shows two exemplary curves of how a network-level metric develops over time for a project with a Code review bot and a matched project without a bot. 
On the x-axis, time points are plotted. 
On the y-axis, the magnitude of a network-level metric is plotted.
Arrows indicate the direction of comparisons.

Time-evolving contribution networks \cite{meneelySociotechnicalDeveloperNetworks2011}, that is, bipartite networks linking developers and edited files, contain rich information on the evolving collective code-ownership in a project at the level of files. Using \textit{git2net}'s feature to construct time-evolving contribution networks, we address the second research question:

\textbf{RQ2:} Does the adoption of Code Review bots influence the structure of \textbf{collective code ownership} in Open Source software projects?

Mirroring the approach for RQ1, we address RQ2 in a two-step process by testing two hypotheses: 

\textbf{H3:} The adoption of Code Review bots is associated with differences in the structure of contribution networks in Open Source projects.

\textbf{H4:} The adoption of Code Review bots in Open Source projects precedes the change of contribution networks.

We will test H3 similarly to how we test H1, except that, different from H1, this requires us to use network metrics that are specifically tailored to bipartite networks, which will be described in \ref{measures}.
For H4, we will use the same approach as for H2, except for the same different network metrics applied.

\section{Data}

\subsection{Repositories}

For the identification of OSS projects hosted on GitHub that use some Code Review bot, we use GitHub Search (GHS) \cite{dabicSamplingProjectsGitHub2021}, a tool to mine GitHub repositories along 25 characteristics such as the number of stars, contributors, or commits.
We determined the availability of data and defined selection criteria for a repository to be included in an initial sample for the preliminary analysis. A repository is included if it:

\begin{itemize}
    \item has at least 3 contributors
    \item has at least 30 commits
    \item is no fork
    \item has its last commit after October 16, 2018 
    \item has at least 10 stars (criterion given by GHS)
    \item has at least an activity of 3 months before and after bot adoption (criterion will be checked after the identification of bot adoption)
\end{itemize}

Extending previous studies, we will also include less popular and smaller repositories in order to achieve higher representativeness and more generalizable results.
In an earlier study, it has been shown that collaboration networks can be constructed for projects of different sizes, even for small teams with only around
ten members\cite{goteBigDataBig2022}. 
We will include repositories with at least three contributors, following the minimal definition of a group \cite{fitziSoziologieUntersuchungenUber2021}, and we will exclude small projects for which collaboration networks cannot be constructed. 
The requirement of a repository having at least 30 commits guarantees that there is a minimum of data available for the analysis of contributions.
We exclude forks, as most of their data is a duplicate of the original repository from which they were forked.
Repositories will be considered if their last commit was made at least after October 16, 2018, the beta-release date of GitHub Actions, and if
their activity data span at least half a year symmetrically around the introduction of a bot.
Data from one month around the introduction of a bot will be excluded from the analysis, as this phase has been observed to be unstable due to the bot introduction \cite{wesselQualityGatekeepersInvestigating2022}.
The inclusion criterion of a repository having at least ten stars is given by GHS.
GHS excludes repositories that have fewer than ten stars to substantially decrease the number of repositories and improve data retrieval.
This does not necessarily ensure that the projects have a certain quality, but it is a good ``compromise between the quality of data and the time required to mine and continuously update all projects" \cite{dabicSamplingProjectsGitHub2021}.

\subsection{GitHub Actions}

Subsequent to the initial selection of projects, we will make use of GitHub's REST API to obtain the contents of a repository's workflow directory if it exists.
The directory \verb+.github/workflows+ usually contains the workflows in the form of .y(a)ml files. 
These .y(a)ml files can then be parsed for the GAs used by a repository. 

\begin{figure}[htbp]
    \centerline{\includegraphics[width=\linewidth]{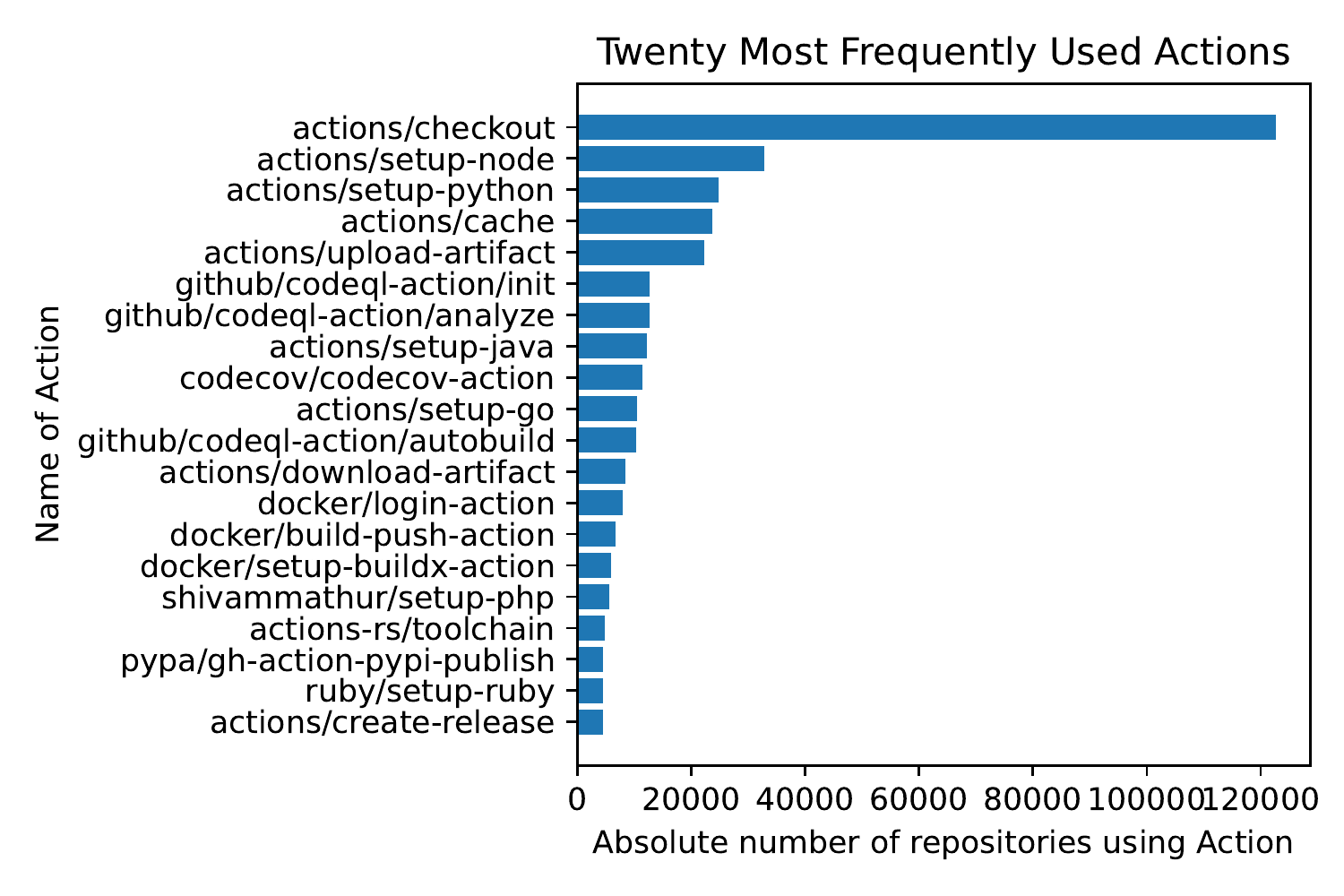}}
    \caption{Twenty most frequently used GitHub Actions used in our sample in descending order. 
    The results stem from a preliminary analysis of about 125,500 repositories that implemented at least one GitHub Action.}
    \label{twenty}
\end{figure}

From GHS, we obtained a list of 319,903 repositories that match the selection criteria described above (except for the activity criterion, which will be considered in a second step), for which we inspected the workflows directory. The list was obtained on January 31, 2023. 
Looking up the repositories and scraping their workflows directory took place between February 6 and 10, 2023.
Our preliminary sampling revealed that about 39\% of the repositories (125,528) use at least one GA, with a minimum of one and a maximum of 63 GAs, and the median being three. 
Fig.~\ref{twenty} gives an overview of the 20 most frequently used GAs across projects, with \verb+actions/checkout+ unsurprisingly being in the first position, used by about 98\% of the  projects.
In the top ten, we also find one GA for code review, \verb+codecov/codecov-action+, used by about 9\% of the projects.

Following the analysis of which GAs are used by the projects, we will finally retrieve the data on the workflow runs and repository history for the repositories with the GAs of interest. 
Based on prior knowledge and exploration of popular Code Review bots, we identified that there are at least the five GAs for code review, listed in Tab.~\ref{codereview}, to serve as GAs of interest. 
Among them is also \verb+Codeball+, a code review Action that uses AI to score and label PRs. 
For \verb+Codecov+, we observed more than ten thousand occurrences.

\begin{table}[htbp]
    \caption{Occurrences of Code Review Actions}
    \begin{center}
        \scalebox{0.87}{
            \begin{tabular}{|p{5cm}|c|c|}
    \hline
    \textbf{GitHub Action} & \textbf{Occurrences} & \textbf{\% of projects} $^{\mathrm{a}}$\\
    \hline
    Codecov \footnote{https://github.com/marketplace/actions/codecov}&  11'373  & 9.0\\
    \hline
    Coveralls GitHub Action\footnote{https://github.com/marketplace/GAs/coveralls-github-action} & 2'311  &  2.0\\
    \hline
    Code Climate Coverage Action \footnote{https://github.com/marketplace/GAs/code-climate-coverage-action}&  479  & \textless 1.0\\
    \hline
    CodeGuru Reviewer \footnote{https://github.com/marketplace/actions/codeguru-reviewer}& 6 &  \textless 1.0\\
    \hline
    Codeball AI Code Review\footnote{https://github.com/marketplace/actions/codeball-ai-code-review} & 28 & \textless 1.0\\
    \hline\hline
    Sum& 14'197 &  \\
    \hline
    \multicolumn{3}{l}{\multilinecell{$^{\mathrm{a}}$Percentages are rounded to one decimal. \\\hspace{0.15em} Repositories may use more than one GA at the same time.}}
    \end{tabular}
    }
    \label{codereview}
    \end{center}
    \end{table}

After obtaining the final selection, we will sample the same number of similar repositories without GAs as a control group.
The similarity will be established in terms of the main programming language, the repository size (number of active contributors, number of commits, number of PRs), and the project age.
This approach allows, for example, to exclude external temporal events as influencing factors.

\section{Execution Plan}

\subsection{Data Retrieval}

We ran an initial script that crawled data through GitHub's REST API to ensure data availability and to get an overview of the number of projects implementing certain GAs.
To identify the usage of GAs, we searched the .y(a)ml files in repositories' workflow directories, if any existed.
As a next step, we will gather data on the events of workflow runs to identify exactly when bots have been introduced and run.
Examining repositories' git histories, we will verify that repositories were active for at least half a year---i.e., that they had contributions--before and after bot adoption in order to be included in the final sample.
We will further verify that projects with Code Review bots did not use another automated code
review solution before switching to Code Review bots based on \textit{GAs}.

\subsection{Statistical Network Analysis}

\subsubsection{General Approach}

Following a socio-technical approach, we will analyze rich, time-stamped data on technical artefacts, communication, and collaboration events that are contained in GitHub repositories.
To obtain data on collaboration networks, we will use the Open Source python software \textit{git2net}, which facilitates the mining of time-stamped co-editing relations between developers from the sequence of ﬁle modiﬁcations contained in git repositories \cite{goteGit2netMiningTimeStamped2019}.
Unlike co-authorship networks, which represent relations between developers and files, co-editing networks allow us to study at great detail how developers edit specific code regions that were previously edited by other developers.
These networks contain rich and fine-grained information on the coordination and collaboration practices of software projects.
With \textit{git2net}, it is possible to construct directed, weighted, and time-stamped networks  \cite{goteGit2netMiningTimeStamped2019}, in particular, 1) co-editing networks linking developers to code written by others and 2) time-stamped bipartite networks linking developers and edited files, which are also known as contribution networks (see also \cite{meneelySociotechnicalDeveloperNetworks2011}).

\subsubsection{Collaboration Networks between Developers (RQ1)}

To answer RQ1, we will adopt the following procedure:

We will first use GHS to generate a representative sample of GitHub repositories. 
We will create two equally-sized, representative samples of GitHub projects, one covering projects that adopted a Code Review bot and a second sample covering projects that did not.
Projects without a bot will be matched based on central characteristics, the project age among others.
To determine whether a project implemented a bot and at which point in time, we will use GitHub's REST API to crawl data on a repository's workflow runs and combine it with information about used GAs, obtained from parsing the .y(a)ml files in the workflow directory.
Using the software package \textit{git2net}, we will extract time-stamped developer co-editing and contribution networks for all projects in both samples.
In a first step, we will use the time-stamped co-editing networks to generate a single time-aggregated weighted co-editing network for each project.
To test hypothesis H1, we will then calculate six network-level metrics for each of the projects in our two samples: 
\begin{itemize}
    \item average node degree (i.e., the average number of co-editing interactions with different developers),
    \item average weighted node degree (average number of total co-editing interactions),
    \item network density (fraction of all possible links that actually exist),
    \item algebraic connectivity, i.e. the second-smallest eigenvalue of the Laplacian matrix, which captures how ``well-connected'' the topology of a social network is and whether it exhibits small cuts or bottlenecks,
    \item average clustering coefficient (average fraction of closed triads around nodes in the network),
    \item degree assortativity (preference of nodes to connect to other nodes with similar degrees).
\end{itemize}
Depending on the distribution of the metrics, we will apply suitable two-sided hypothesis tests to respectively test the null hypothesis that a network metric is the same for those projects that do not use Code Review bots and those that do use Code Review bots, against the alternative hypothesis that there is a statistically significant difference.

For H2, we use \textit{git2net} to generate \textit{time-stamped} co-editing networks. 
Based on the date of the first introduction of a Code Review bot in a project, we will split the project data into two phases where each phase covers a time interval of six months:
A first phase \textit{before} the introduction of the bot and a second phase \textit{after} its introduction.
Reusing the network metrics outlined above, we will generate time-aggregated networks for each of these two phases individually, obtaining---for each project---a co-editing network before and after the introduction of a Code Review bot.
This allows us to apply paired two-sample tests to test the null hypothesis that there is no significant change in the topology of co-editing networks against the alternative hypothesis that there is a significant change.
To control for temporal effects such as effects evoked by project ageing, we will conduct multiple additional intra-project comparisons. With this, we can determine the magnitude of ``regular'' changes in  network metrics and evaluate whether the changes observed for the bot adoption significantly deviate from the distribution of changes observed for other time points.
In an effort to avoid a large sample size bias \cite{kaplanBigDataLarge2014}, we will make sure to use a significance threshold that is appropriate given the underlying sample sizes. 

\subsubsection{Collective Code Ownership Networks (RQ2)}
\label{measures}
To address RQ2, we construct time-evolving bipartite networks linking developers and edited files, also known as contribution networks (see also \cite{meneelySociotechnicalDeveloperNetworks2011}).
These networks contain rich socio-technical information on the evolving collective code-ownership in a project at the level of files.
To answer RQ2, we use the same approach as for RQ1, except that we need to apply different network metrics, specifically tailored to bipartite networks.
In particular, we can calculate the degree-based metrics for both partitions (developers and files) of the contributions separately. 
For computing the network density, we will need a definition that accounts for the maximum number of links in a bipartite network (which is smaller than in a general graph with a single partition).
Since the metrics algebraic connectivity, clustering coefficient, and degree assortativity are not meaningful in a bipartite network, we will additionally calculate the number of connected components in a one-mode projection of the bipartite network, where we project the bipartite network to files.

\subsection{Qualitative Analysis}

We aim to qualitatively interpret the changes in the collaboration structure we observe in our exploratory study. To this end, we plan to present our findings to developers from the corresponding software development communities and ask for their interpretation.

\section{Limitations}

In order to identify projects utilizing Code Review bots, in our preliminary analysis we checked the repositories' workflow directory for occurrences of GAs at the time of the sampling.
However, it could be the case that repositories previously implemented a GA but later removed it, with the consequence that there would be no trace of it in the \verb+.github/workflows+ directory at the time of the sampling.
However, the combination of data on the time stamps of workflow runs and the commit log history of the workflow directory should also enable us to identify projects that used bots in the past.
We will adopt this approach in the execution of the proposed research.

As a cautionary note, we highlight that while our aim is to study the effect of bots on collaboration, we focus on the examination of interaction patterns that emerge in code collaboration practices. 
With that, we cannot capture other forms of interaction, such as exchange happening in external communication channels. 

\section{Summary}

This registered report proposes a novel approach to examine the impact of Code Review bots on Open Source Software engineering teams on GitHub.
Taking a network-analytic perspective, we propose to study the effect of bot adoption on such teams by examining changes in the collaboration structures of development team.
We strongly believe that we need a firm understanding of this collective dimension of the effect of bots on development and collaboration practices, and that we currently lack insights into potential unintended consequences of bot adoption.
To address this gap, we will retrieve time-stamped co-editing and developer-file contribution networks and compare suitable network metrics between a sample of comparable projects that adopt bots and projects that do not.
We will further make a comparison of network metrics within projects that adopt a bot computed for the phase before and after bot adoption. 
With this approach, and relying on a larger and more representative sample of GitHub repositories that also includes less popular projects, we aim to advance our knowledge on how software bots in general, and Code Review bots in particular, impact the way members of Open Source software development teams collaborate. 
This may also help to better understand human-bot interaction and improve the ways in which intelligent tools can support teams, in general.

\section*{Acknowledgment}
We gratefully acknowledge the financial support from Honda Research Institute Europe (HRI-EU).
I.S. and C.G. acknowledge support by the Swiss National Science Foundation, grant 176938.
The authors thank Johannes Wachs for inspiration in the ideation phase of the project.

\bibliography{chase23_clean}

% Generated by IEEEtran.bst, version: 1.14 (2015/08/26)
\begin{thebibliography}{10}
\providecommand{\url}[1]{#1}
\csname url@samestyle\endcsname
\providecommand{\newblock}{\relax}
\providecommand{\bibinfo}[2]{#2}
\providecommand{\BIBentrySTDinterwordspacing}{\spaceskip=0pt\relax}
\providecommand{\BIBentryALTinterwordstretchfactor}{4}
\providecommand{\BIBentryALTinterwordspacing}{\spaceskip=\fontdimen2\font plus
\BIBentryALTinterwordstretchfactor\fontdimen3\font minus
  \fontdimen4\font\relax}
\providecommand{\BIBforeignlanguage}[2]{{%
\expandafter\ifx\csname l@#1\endcsname\relax
\typeout{** WARNING: IEEEtran.bst: No hyphenation pattern has been}%
\typeout{** loaded for the language `#1'. Using the pattern for}%
\typeout{** the default language instead.}%
\else
\language=\csname l@#1\endcsname
\fi
#2}}
\providecommand{\BIBdecl}{\relax}
\BIBdecl

\bibitem{dabbishSocialCodingGitHub2012}
L.~Dabbish, C.~Stuart, J.~Tsay, and J.~Herbsleb, ``Social {{Coding}} in
  {{GitHub}}: {{Transparency}} and {{Collaboration}} in an {{Open Software
  Repository}},'' in \emph{Proceedings of the {{ACM}} 2012 Conference on
  {{Computer Supported Cooperative Work}}}.\hskip 1em plus 0.5em minus
  0.4em\relax {Seattle Washington USA}: {ACM}, Feb. 2012, pp. 1277--1286.

\bibitem{schunemannSocialCodingRevolution2018}
D.~Sch{\"u}nemann, \emph{Die {{Social-Coding-Revolution}}}.\hskip 1em plus
  0.5em minus 0.4em\relax {Wiesbaden}: {Springer Fachmedien Wiesbaden}, 2018.

\bibitem{lebeufDefiningClassifyingSoftware2019}
C.~Lebeuf, A.~Zagalsky, M.~Foucault, and M.-A. Storey, ``Defining and
  {{Classifying Software Bots}}: {{A Faceted Taxonomy}},'' in \emph{2019
  {{IEEE}}/{{ACM}} 1st {{International Workshop}} on {{Bots}} in {{Software
  Engineering}} ({{BotSE}})}.\hskip 1em plus 0.5em minus 0.4em\relax {Montreal,
  QC, Canada}: {IEEE}, May 2019, pp. 1--6.

\bibitem{GitHubActionsFocus2023}
``{{GitHub Actions}}: {{Focus}} on what matters: Code. {{GitHub}},'' in
  \emph{{{https://web.archive.org/web/20181016171717/https://github.com/features}}
  /Actions}, (accessed Feb. 14, 2023).

\bibitem{GitHubActionsGenerally2023}
``{{GitHub Actions}} is generally available. {{GitHub Blog}},'' in
  \emph{{{https://github.blog/changelog/2019-11-11-github-actions-is-generally-available/}}},
  (accessed April. 13, 2023).

\bibitem{erlenhovCurrentFutureBots2019}
L.~Erlenhov, F.~G. {de Oliveira Neto}, R.~Scandariato, and P.~Leitner,
  ``Current and {{Future Bots}} in {{Software Development}},'' in \emph{2019
  {{IEEE}}/{{ACM}} 1st {{International Workshop}} on {{Bots}} in {{Software
  Engineering}} ({{BotSE}})}.\hskip 1em plus 0.5em minus 0.4em\relax {IEEE},
  2019, pp. 7--11.

\bibitem{storeyDisruptingDeveloperProductivity2016}
M.-A. Storey and A.~Zagalsky, ``Disrupting {{Developer Productivity One Bot}}
  at a {{Time}},'' in \emph{Proceedings of the 2016 24th {{ACM SIGSOFT
  International Symposium}} on {{Foundations}} of {{Software
  Engineering}}}.\hskip 1em plus 0.5em minus 0.4em\relax {Seattle WA USA}:
  {ACM}, Nov. 2016, pp. 928--931.

\bibitem{zampettiEmpiricalCharacterizationBad2020}
F.~Zampetti, C.~Vassallo, S.~Panichella, G.~Canfora, H.~Gall, and M.~Di~Penta,
  ``An {{Empirical Characterization}} of {{Bad Practices}} in {{Continuous
  Integration}},'' \emph{Empirical Software Engineering}, vol.~25, no.~2, pp.
  1095--1135, Mar. 2020.

\bibitem{rueStateSoftwareCode2021}
B.~Rue, ``The {{State}} of {{Software Code Report}} 2021,'' \emph{The State of
  Software Code Report 2021}, p.~11, 2021.

\bibitem{huImprovingFeedbackGitHub2019}
Z.~Hu and E.~F. Gehringer, ``Improving {{Feedback}} on {{GitHub Pull
  Requests}}: {{A Bots Approach}},'' in \emph{2019 {{IEEE Frontiers}} in
  {{Education Conference}} ({{FIE}})}.\hskip 1em plus 0.5em minus 0.4em\relax
  {Covington, KY, USA}: {IEEE}, Oct. 2019, pp. 1--9.

\bibitem{kinsmanHowSoftwareDevelopers2021a}
T.~Kinsman, M.~Wessel, M.~A. Gerosa, and C.~Treude, ``How {{Do Software
  Developers Use GitHub Actions}} to {{Automate Their Workflows}}?''\hskip 1em
  plus 0.5em minus 0.4em\relax {arXiv preprint arXiv.2103.12224}, Mar. 2021.

\bibitem{wesselPowerBotsUnderstanding2018}
M.~Wessel, B.~M. {de Souza}, I.~Steinmacher, I.~S. Wiese, I.~Polato, A.~P.
  Chaves, and M.~A. Gerosa, ``The {{Power}} of {{Bots}}: {{Understanding Bots}}
  in {{OSS Projects}},'' \emph{Proceedings of the ACM on Human-Computer
  Interaction}, vol.~2, no. CSCW, pp. 1--19, Nov. 2018.

\bibitem{wesselQualityGatekeepersInvestigating2022}
M.~Wessel, A.~Serebrenik, I.~Wiese, I.~Steinmacher, and M.~A. Gerosa, ``Quality
  {{Gatekeepers}}: {{Investigating}} the {{Effects}} of {{Code Review Bots}} on
  {{Pull Request Activities}},'' \emph{Empirical Software Engineering},
  vol.~27, no.~5, p. 108, Sep. 2022.

\bibitem{farahExploratoryStudyReactions2022a}
J.~C. Farah, B.~Spaenlehauer, X.~Lu, S.~Ingram, and D.~Gillet, ``An
  {{Exploratory Study}} of {{Reactions}} to {{Bot Comments}} on {{GitHub}},''
  in \emph{4th {{International Workshop}} on {{Bots}} in {{Software
  Engineering}} ({{BotSE}} 2022)}, no. CONF.\hskip 1em plus 0.5em minus
  0.4em\relax {ACM}, 2022.

\bibitem{CodeballAICode2023}
``Codeball {{AI Code Review}}. {{GitHub}},'' in
  \emph{{{https://github.com/marketplace/actions/codeball-ai-code-review}}},
  (accessed Feb. 14, 2023).

\bibitem{wesselBotsPullRequests2022}
M.~Wessel, A.~Abdellatif, I.~Wiese, T.~Conte, E.~Shihab, M.~A. Gerosa, and
  I.~Steinmacher, ``Bots for {{Pull Requests}}: {{The Good}}, the {{Bad}}, and
  the {{Promising}},'' in \emph{Proceedings of the 44th {{International
  Conference}} on {{Software Engineering}}}.\hskip 1em plus 0.5em minus
  0.4em\relax {Pittsburgh Pennsylvania}: {ACM}, May 2022, pp. 274--286.

\bibitem{deyImpactTeamDesign2020}
C.~Dey and G.~M.P., ``Impact of {{Team Design}} and {{Technical Factors}} on
  {{Team Cohesion}},'' \emph{Team Performance Management: An International
  Journal}, vol.~26, no. 7/8, pp. 357--374, Aug. 2020.

\bibitem{wesselGitHubActionsImpact2022}
M.~Wessel, J.~Vargovich, M.~A. Gerosa, and C.~Treude, ``{{GitHub Actions}}:
  {{The Impact}} on the {{Pull Request Process}},'' Jun. 2022.

\bibitem{chenLetSuperchargeWorkflows2021}
T.~Chen, Y.~Zhang, S.~Chen, T.~Wang, and Y.~Wu, ``Let's {{Supercharge}} the
  {{Workflows}}: {{An Empirical Study}} of {{GitHub Actions}},'' in \emph{2021
  {{IEEE}} 21st {{International Conference}} on {{Software Quality}},
  {{Reliability}} and {{Security Companion}} ({{QRS-C}})}.\hskip 1em plus 0.5em
  minus 0.4em\relax {Hainan, China}: {IEEE}, Dec. 2021, pp. 01--10.

\bibitem{goteBigDataBig2022}
C.~Gote, P.~Mavrodiev, F.~Schweitzer, and I.~Scholtes, ``Big data = big
  insights?: {{Operationalising Brooks}}' {{Law}} in a {{Massive GitHub Data
  Set}},'' in \emph{Proceedings of the 44th {{International Conference}} on
  {{Software Engineering}}}.\hskip 1em plus 0.5em minus 0.4em\relax {Pittsburgh
  Pennsylvania}: {ACM}, May 2022, pp. 262--273.

\bibitem{golzadehRiseFallCI2022}
M.~Golzadeh, A.~Decan, and T.~Mens, ``On the {{Rise}} and {{Fall}} of {{CI
  Services}} in {{GitHub}},'' in \emph{2022 {{IEEE International Conference}}
  on {{Software Analysis}}, {{Evolution}} and {{Reengineering}}
  ({{SANER}})}.\hskip 1em plus 0.5em minus 0.4em\relax {Honolulu, HI, USA}:
  {IEEE}, Mar. 2022, pp. 662--672.

\bibitem{zhaoImpactContinuousIntegration2017}
Y.~Zhao, A.~Serebrenik, Y.~Zhou, V.~Filkov, and B.~Vasilescu, ``The {{Impact}}
  of {{Continuous Integration}} on other {{Software Development Practices}}:
  {{A Large-Scale Empirical Study}},'' in \emph{2017 32nd {{IEEE}}/{{ACM
  International Conference}} on {{Automated Software Engineering}}
  ({{ASE}})}.\hskip 1em plus 0.5em minus 0.4em\relax {Urbana, IL}: {IEEE}, Oct.
  2017, pp. 60--71.

\bibitem{moharilJIRAGitHubASFBot2022}
A.~Moharil, D.~Orlov, S.~Jameel, T.~Trouwen, N.~Cassee, and A.~Serebrenik,
  ``Between {{JIRA}} and {{GitHub}}: {{ASFBot}} and its {{Influence}} on
  {{Human Comments}} in {{Issue Trackers}},'' in \emph{19th {{International
  Conference}} on {{Mining SoftwareRepositories}} ({{MSR}} '22)}, {Pittsburgh,
  PA, USA}, 2022, p.~5.

\bibitem{mohayejiAdoptionTODOBot2022}
H.~Mohayeji, F.~Ebert, and E.~Arts, ``On the {{Adoption}} of a {{TODO Bot}} on
  {{GitHub}}: {{A Preliminary Study}},'' in \emph{4th {{International
  Workshop}} on {{Bots}} in {{Software Engineering}} ({{BotSE}} 2022)},
  {Pittsburgh, PA, USA}, 2022, p.~5.

\bibitem{casseeSilentHelperImpact2020}
N.~Cassee, B.~Vasilescu, and A.~Serebrenik, ``The {{Silent Helper}}: {{The
  Impact}} of {{Continuous Integration}} on {{Code Reviews}},'' in \emph{2020
  {{IEEE}} 27th {{International Conference}} on {{Software Analysis}},
  {{Evolution}} and {{Reengineering}} ({{SANER}})}.\hskip 1em plus 0.5em minus
  0.4em\relax {IEEE}, 2020, pp. 423--434.

\bibitem{guoStudyingImpactCI2019}
Y.~Guo and P.~Leitner, ``Studying the {{Impact}} of {{CI}} on {{Pull Request
  Delivery Time}} in {{Open Source Projects}}\textemdash{{A Conceptual
  Replication}},'' \emph{PeerJ Computer Science}, vol.~5, p. e245, 2019.

\bibitem{zimmermannImpactSwitchingBug2019}
T.~Zimmermann and A.~C. Art{\'i}s, ``Impact of {{Switching Bug Trackers}}: {{A
  Case Study}} on a {{Medium-Sized Open Source Project}},'' in \emph{2019 Ieee
  International Conference on Software Maintenance and Evolution
  (Icsme)}.\hskip 1em plus 0.5em minus 0.4em\relax {IEEE}, 2019, pp. 13--23.

\bibitem{birdDonTouchMy2011}
C.~Bird, N.~Nagappan, B.~Murphy, H.~Gall, and P.~Devanbu, ``Don't touch my
  code!: {{Examining}} the {{Effects}} of {{Ownership}} on {{Software
  Quality}},'' in \emph{Proceedings of the 19th {{ACM SIGSOFT}} Symposium and
  the 13th {{European}} Conference on {{Foundations}} of Software
  Engineering}.\hskip 1em plus 0.5em minus 0.4em\relax {Szeged Hungary}: {ACM},
  Sep. 2011, pp. 4--14.

\bibitem{birdPuttingItAll2009a}
C.~Bird, N.~Nagappan, H.~Gall, B.~Murphy, and P.~Devanbu, ``Putting {{It All
  Together}}: {{Using Socio-technical Networks}} to {{Predict Failures}},'' in
  \emph{2009 20th {{International Symposium}} on {{Software Reliability
  Engineering}}}.\hskip 1em plus 0.5em minus 0.4em\relax {Mysuru, Karnataka,
  India}: {IEEE}, Nov. 2009, pp. 109--119.

\bibitem{joblinHowSuccessfulFailed2022}
M.~Joblin and S.~Apel, ``How {{Do Successful}} and {{Failed Projects Differ}}?
  {{A Socio-Technical Analysis}},'' \emph{ACM Transactions on Software
  Engineering and Methodology}, vol.~31, no.~4, pp. 1--24, Oct. 2022.

\bibitem{joblinDeveloperNetworksVerified2015}
M.~Joblin, W.~Mauerer, S.~Apel, J.~Siegmund, and D.~Riehle, ``From {{Developer
  Networks}} to {{Verified Communities}}: {{A Fine-Grained Approach}},'' in
  \emph{2015 {{IEEE}}/{{ACM}} 37th {{IEEE International Conference}} on
  {{Software Engineering}}}.\hskip 1em plus 0.5em minus 0.4em\relax {Florence,
  Italy}: {IEEE}, May 2015, pp. 563--573.

\bibitem{meneelySociotechnicalDeveloperNetworks2011}
A.~Meneely and L.~Williams, ``Socio-technical {{Developer Networks}}: {{Should
  We Trust Our Measurements}}?'' in \emph{Proceedings of the 33rd
  {{International Conference}} on {{Software Engineering}}}.\hskip 1em plus
  0.5em minus 0.4em\relax {Waikiki, Honolulu HI USA}: {ACM}, May 2011, pp.
  281--290.

\bibitem{lopez-fernandezApplyingSocialNetwork2006}
L.~{L{\'o}pez-Fern{\'a}ndez}, G.~Robles, J.~M. {Gonzalez-Barahona}, and
  I.~Herraiz, ``Applying {{Social Network Analysis Techniques}} to
  {{Community-Driven Libre Software Projects}}:,'' \emph{International Journal
  of Information Technology and Web Engineering}, vol.~1, no.~3, pp. 27--48,
  Jul. 2006.

\bibitem{goteGit2netMiningTimeStamped2019}
C.~Gote, I.~Scholtes, and F.~Schweitzer, ``Git2net - {{Mining Time-Stamped
  Co-Editing Networks}} from {{Large}} git {{Repositories}},'' \emph{2019
  IEEE/ACM 16th International Conference on Mining Software Repositories
  (MSR)}, pp. 433--444, May 2019.

\bibitem{trujilloCorrectionPenumbraOpen2022}
M.~Z. Trujillo, L.~{H{\'e}bert-Dufresne}, and J.~Bagrow, ``Correction: {{The
  Penumbra}} of {{Open Source}}: {{Projects Outside}} of {{Centralized
  Platforms}} are {{Longer Maintained}}, {{More Academic}} and {{More
  Collaborative}},'' \emph{EPJ Data Science}, vol.~11, no.~1, p.~37, Dec. 2022.

\bibitem{goteAnalysingTimeStampedCoEditing2021a}
C.~Gote, I.~Scholtes, and F.~Schweitzer, ``Analysing {{Time-Stamped Co-Editing
  Networks}} in {{Software Development Teams}} using git2net,'' \emph{Empirical
  Software Engineering}, vol.~26, no.~4, p.~75, Jul. 2021.

\bibitem{scholtesAristotleRingelmannLargeScale2016}
I.~Scholtes, P.~Mavrodiev, and F.~Schweitzer, ``From {{Aristotle}} to
  {{Ringelmann}}: A {{Large-Scale Analysis}} of {{Team Productivity}} and
  {{Coordination}} in {{Open Source Software Projects}},'' \emph{Empirical
  Software Engineering}, vol.~21, no.~2, pp. 642--683, Apr. 2016.

\bibitem{dabicSamplingProjectsGitHub2021}
O.~Dabic, E.~Aghajani, and G.~Bavota, ``Sampling {{Projects}} in {{GitHub}} for
  {{MSR Studies}},'' in \emph{2021 {{IEEE}}/{{ACM}} 18th {{International
  Conference}} on {{Mining Software Repositories}} ({{MSR}})}.\hskip 1em plus
  0.5em minus 0.4em\relax {IEEE}, Mar. 2021, pp. 560--564.

\bibitem{fitziSoziologieUntersuchungenUber2021}
G.~Fitzi, ``Soziologie. {{Untersuchungen}} \"uber die {{Formen}} der
  {{Vergesellschaftung}} (1908),'' \emph{Simmel-Handbuch: Leben\textendash
  Werk\textendash Wirkung}, pp. 235--244, 2021.

\bibitem{kaplanBigDataLarge2014}
R.~M. Kaplan, D.~A. Chambers, and R.~E. Glasgow, ``Big {{Data}} and {{Large
  Sample Size}}: {{A Cautionary Note}} on the {{Potential}} for {{Bias}},''
  \emph{Clinical and Translational Science}, vol.~7, no.~4, pp. 342--346, Aug.
  2014.

\end{thebibliography}
\bibliographystyle{IEEEtran}

\end{document}